\documentstyle[aps,epsf,preprint]{revtex}
\tightenlines
\begin{document}
\draft
\title{The decay $\omega(782)\to5\pi$ in chiral theory.}
\author{N.~N.~Achasov \footnote{Electronic address: achasov@math.nsc.ru}
and
A.~A.~Kozhevnikov \footnote{Electronic address: kozhev@math.nsc.ru}}
\address{Laboratory of Theoretical Physics, \\
Sobolev Institute for Mathematics \\
630090, Novosibirsk-90, Russia}
\date{\today}
\maketitle
\begin{abstract}
The arguments are put forward that the many pion decays
$\omega\to2\pi^+2\pi^-\pi^0$ and $\pi^+\pi^-3\pi^0$ provide an
ideal test site for testing the predictions of chiral models of
the vector meson decays into many pions. Using the approach based
on the Weinberg Lagrangian or, in a new language, the Lagrangian
of hidden local symmetry added with the term induced by the
anomalous Lagrangian of Wess and Zumino, the partial widths of
these decays are evaluated, and their excitation curves in
$e^+e^-$ annihilation are obtained. The discussed are the
perspectives of the experimental study of the decays
$\omega\to5\pi$ in $e^+e^-$ annihilation and photoproduction.
\end{abstract}
\narrowtext

At present, the unconventional, from the point of view of the chiral pion
dynamics, sources of soft pions are feasible. Indeed, the progress in
increasing the luminosity of low energy $e^+e^-$ colliders ($\phi$ factories)
\cite{phifac} could offer the
naturally controlled sources of soft pions. Other possible source of such pions
could be the intense  photon beams \cite{dzierba} , provided the sufficiently
low invariant mass regions of the many pion systems are isolated.
The yield of pions is considerably enhanced when they are produced through
proper vector resonance states. Then, choosing the many pion decays of
sufficiently low lying resonances, one can obtain the soft pions in
quantities sufficient for testing the predictions of chiral models that
include vector mesons.

The decay $\omega(782)\to5\pi$ whose final state pions possess the
momenta $|{\bf q}_\pi|\simeq 74$ MeV, is just of this kind. The
latter value is sufficiently small to expect the manifestation of
chiral dynamics in most clean form. By this we mean that the
higher derivative and loop terms in the effective Lagrangian are
severely suppressed. The present paper is devoted to the
evaluation of the partial width of this decay and plotting its
excitation curve in $e^+e^-$ annihilation.

The $\rho\pi$ sector is considered here on the basis of the
Weinberg Lagrangian \cite{weinberg68} revived later as the
Lagrangian of hidden local symmetry (HLS) \cite{bando}. The former
looks as
\begin{eqnarray}
{\cal L}&=&-{1\over4}\left(\partial_\mu\bbox{\rho}_\nu-
\partial_\nu\bbox{\rho}_\mu+g[\bbox{\rho}_\mu\times\bbox{\rho}_\nu]
\right)^2     \nonumber\\
&&+{m^2_\rho\over2}\left[\bbox{\rho}_\mu+
{[\bbox{\pi}\times\partial_\mu\bbox{\pi}]\over2g f^2_\pi
(1+\bbox{\pi}^2/4f^2_\pi)}\right]^2      \nonumber\\
&&+{(\partial_\mu\bbox{\pi})^2\over2\left(1+\bbox{\pi}^2
/4f^2_\pi\right)^2}
-{m^2_\pi\bbox{\pi}^2\over2(1+\bbox{\pi}^2/4f^2_\pi)},
\label{lwein}
\end{eqnarray}
where $\bbox{\pi}, m_\pi$ and $\bbox{\rho}, m_\rho$ stand for the
isovector field and mass of pion and $\rho$ meson, respectively,
cross denotes the vector product in the isovector space, and
$f_\pi=92.4$ MeV is the pion decay constant. The $\rho\rho\rho$
coupling constant $g$ and the $\rho\pi\pi$ coupling constant
$g_{\rho\pi\pi}$ are related to the $\rho$ mass and pion decay
constant $f_\pi$ via the parameter of hidden local symmetry $a$ as
\cite{bando}
\begin{eqnarray}
g&=&m_\rho/f_\pi\sqrt{a},   \nonumber\\
g_{\rho\pi\pi}&=&\sqrt{a}m_\rho/2f_\pi.
\label{a}
\end{eqnarray}
Note that  $a=2$,
if one demands the universality condition $g=g_{\rho\pi\pi}$
to be satisfied. Then the so called  Kawarabayashi-Suzuki-Riazzuddin-
Fayyazuddin (KSRF) relation \cite{ksrf} arises
\begin{equation}
2g^2_{\rho\pi\pi}f^2_\pi/m^2_\rho=1
\label{ksrf}
\end{equation}
which beautifully agrees with experiment. The $\rho\pi\pi$
coupling constant resulting from this relation is
$g_{\rho\pi\pi}=5.89$. The inclusion of the interaction of the
$\omega(782)$ with the $\rho\pi$ state is achieved upon adding
the term induced by the anomalous Lagrangian of Wess and Zumino
\cite{bando,wza},
\begin{equation}
{\cal L}_{\omega\rho\pi}={N_cg^2\over8\pi^2 f_\pi}
\varepsilon_{\mu\nu\lambda\sigma}\partial_\mu\omega_\nu\left(\bbox{\pi}
\cdot\partial_\lambda\bbox{\rho}_\sigma\right),
\label{wz}
\end{equation}
where $N_c=3$ is the number of colors, $\omega_\nu$ is the field
of $\omega$ meson.

\begin{figure}
\centerline {\epsfysize=7in \epsfbox{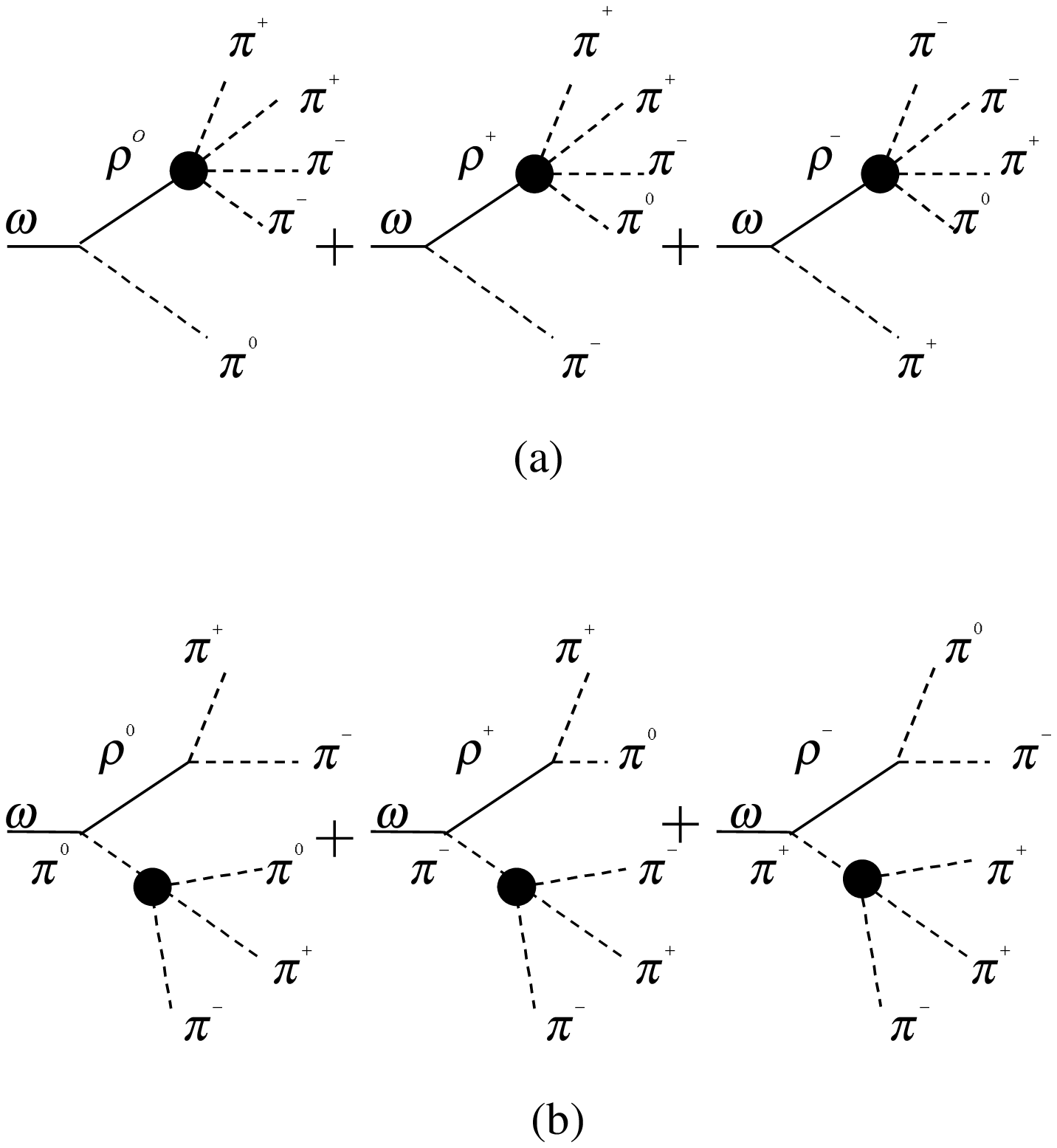}}
\caption{The diagrams describing the amplitudes of the decay
$\omega\to\pi^+\pi^-\pi^+\pi^-\pi^0$.
The shaded circles in the set (a) refer to the whole
$\rho\to4\pi$ amplitudes Eq.~(\protect\ref{nonrel}).
The shaded circles in the set (b)  refer to the
effective $\pi\to3\pi$ vertices given by Eq.~(\protect\ref{4pinr}).
The symmetrisation over  momenta of
identical pions is meant.
The diagrams for the decay $\omega\to\pi^+\pi^-\pi^0\pi^0\pi^0$ are obtained
from those shown  upon the evident replacements.
\label{fig1}}
\end{figure}

One may convince oneself that the $\omega\to\rho\pi\to5\pi$ decay
amplitude unambiguously results from the Weinberg Lagrangian
Eq.~(\ref{lwein}) and the anomaly induced Lagrangian (\ref{wz}).
This amplitude is represented by the diagrams shown in
Fig.~\ref{fig1}. As one can foresee, its general expression looks
cumbersome. However, it can be considerably simplified upon noting
that the small pion momenta permit one to use the nonrelativistic
expressions,
\begin{eqnarray}
M(\rho^0\to\pi^+_{q_1}\pi^+_{q_2}\pi^-_{q_3}\pi^-_{q_4})
&\simeq&-{g_{\rho\pi\pi}\over2f^2_\pi}
(\varepsilon,q_1+q_2-q_3-q_4),      \nonumber\\
M(\rho^0\to\pi^+_{q_1}\pi^-_{q_2}\pi^0_{q_3}\pi^0_{q_4})
&\simeq&-{g_{\rho\pi\pi}\over4f^2_\pi}
(\varepsilon,q_1-q_2),      \nonumber\\
M(\rho^+\to\pi^+_{q_1}\pi^+_{q_2}\pi^-_{q_3}\pi^0_{q_4})
&\simeq&{g_{\rho\pi\pi}\over4f^2_\pi}
(\varepsilon,q_1+q_2-2q_4),      \nonumber\\
M(\rho^+\to\pi^+_{q_1}\pi^0_{q_2}\pi^0_{q_3}\pi^0_{q_4})
&\simeq&{g_{\rho\pi\pi}\over f^2_\pi}
(\varepsilon,q_1),
\label{nonrel}
\end{eqnarray}
for the $\rho\to4\pi$ decay amplitudes in the diagrams
Fig.~\ref{fig1}(a), where $\varepsilon$ stands for the $\rho$
meson polarization four-vector \footnote{The notation for the
Lorentz invariant product of two four-vectors $a$ and $b$ is
$(a,b)=a_0b_o-{\bf a}{\bf b}$.} . They are valid with the accuracy
$5\%$ in the $4\pi$ mass range relevant for the present purpose.
Likewise, the expression for the combination $D^{-1}_\pi
M(\pi\to3\pi)$ standing in the expression for the diagrams in
Fig.~\ref{fig1}(b) can be replaced, with the same accuracy, by
$-(8m^2_\pi)^{-1}$ times the nonrelativistic $\pi\to3\pi$
amplitudes. The latter look as
\begin{eqnarray}
M(\pi^+\to\pi^+\pi^+\pi^-)&=&-2m^2_\pi/f^2_\pi, \nonumber\\
M(\pi^+\to\pi^+\pi^0\pi^0)&=&-m^2_\pi/f^2_\pi, \nonumber\\
M(\pi^0\to\pi^+\pi^-\pi^0)&=&-m^2_\pi/f^2_\pi, \nonumber\\
M(\pi^0\to\pi^0\pi^0\pi^0)&=&-3m^2_\pi/f^2_\pi. \label{4pinr}
\end{eqnarray}
Note that, in the nonrelativistic limit, the $\rho\to4\pi$ decay
amplitudes depend on the HLS parameter $a$ only through an overall
factor $g_{\rho\pi\pi}/f^2_\pi=\sqrt{a}m_\rho/2f^3_\pi,$ while the
$\pi\to3\pi$ amplitudes do not depend on it at all.

The LHS approach \cite{bando} permits one to include the axial
mesons as well \footnote{The problem of inclusion of vector, axial
mesons, and photons in the framework of chiral theories has
demanded considerable efforts. See \cite{efforts}. It is solved in
an elegant way in the HLS approach \cite{bando}.}.  An ideal
treatment would consist of that under the assumption of
$m_\rho\sim E \ll m_{a_1}$, the difference between the models with
and without $a_1$ meson would be reduced to the allowing for the
higher derivatives and would be small \footnote{Allowing for the
higher derivatives demands also allowing for the chiral loops, the
task which is not yet fulfilled for vector mesons.}. In
particular, the correction to the $\rho^0\to4\pi$ decay width due
to $a_1$ meson is estimated at the level of 30-40 $\%$
\cite{birse}. However, since the invariant mass of four pions in
the diagrams Fig.~\ref{fig1}(a) is less than 642 MeV, the $a_1$
contribution in the $\omega\to5\pi$ decay is severely suppressed.
Furthermore, the introducing $a_1$ meson results in the terms
$\propto(\partial_\mu\bbox{\pi})^4$ in the Lagrangian, but their
contribution to the $\pi\to3\pi$ transition amplitude is
inessential in the relevant three pion invariant mass region less
than 500 MeV.

Note also that  the maximum pion momentum $|{\bf q}_\pi|$ is about
115 MeV, but the contribution of such regions of the phase space
is negligible. The dominant contribution comes from the momentum
$|{\bf q}_\pi|\approx 70$ MeV. In the kinematical situation of the
$\omega\to5\pi$ decay we also neglect  the contributions of other
higher derivatives and quantum corrections due to chiral loops.
Thus, our approach is zeroth order approximation to the
$\omega\to5\pi$ decay amplitude in the framework of the
Lagrangians (\ref{lwein}) and (\ref{wz}), in a close analogy with
the Weinberg amplitudes in classical $\pi\pi$ scattering. In this
approximation all chiral models of the vector meson interactions
with pions  indistinguishable, and any difference could manifest
upon including the higher derivatives and chiral loops.

Then one obtains, upon neglecting the corrections of the order of
$O(|{\bf q}_\pi|^4/m^4_\pi)$ or higher, the expression for decay
amplitudes: \widetext
\begin{eqnarray}
M(\omega\to2\pi^+2\pi^-\pi^0)&=&{N_cg_{\rho\pi\pi}g^2\over8(2\pi)^2
f^3_\pi}\varepsilon_{\mu\nu\lambda\sigma}q_\mu\epsilon_\nu
\left\{(1+P_{12})q_{1\lambda}\left[{(q_2+3q_4)_\sigma\over
D_\rho(q-q_1)} -{2q_{4\sigma}\over D_\rho(q_1+q_4)}\right]\right.
\nonumber\\
&&\left.-(1+P_{35})q_{3\lambda}\left[{(q_5+3q_4)_\sigma\over
D_\rho(q-q_3)} -{2q_{4\sigma}\over D_\rho(q_3+q_4)}\right]\right.
\nonumber\\ &&\left.-
(1+P_{12})(1+P_{35})q_{3\lambda}\left[{2q_{4_\sigma}\over
D_\rho(q-q_4)} +{q_{1\sigma}\over D_\rho(q_1+q_3)}\right]\right\},
\label{om1pi0}
\end{eqnarray}
\narrowtext
with the final momentum  assignment according to
$\pi^+(q_1)\pi^+(q_2)\pi^-(q_3)\pi^-(q_5)\pi^0(q_4)$, and
\widetext
\begin{eqnarray}
M(\omega\to\pi^+\pi^-3\pi^0)&=&{N_cg_{\rho\pi\pi}g^2\over8(2\pi)^2
f^3_\pi}(1-P_{12})(1+P_{34}+P_{35})
\varepsilon_{\mu\nu\lambda\sigma}q_\mu\epsilon_\nu q_{1\lambda}
\nonumber\\ &&\times\left\{q_{3\sigma}\left[{1\over D_\rho(q-q_3)}
-{1\over D_\rho(q_1+q_3)}\right]\right.\nonumber\\
&&\left.-q_{2\sigma}\left[{4\over3D_\rho(q-q_1)} -{1\over
2D_\rho(q_1+q_2)}\right]\right\},
\label{om3pi0}
\end{eqnarray}
\narrowtext
with the final momentum  assignment according to
$\pi^+(q_1)\pi^-(q_2)\pi^0(q_3)\pi^0(q_4)\pi^0(q_5)$.
Note that the first term
in each square bracket refers to the specific diagram
shown in Fig.~\ref{fig1}(a) while the second one does to the diagram
shown in Fig.~\ref{fig1}(b).
In both above formulas, $\epsilon_\nu$, $q_\mu$ stand for four-vectors
of polarization and momentum of $\omega$ meson, $P_{ij}$ is the operator
of the interchange the pion momenta $q_i$ and $q_j$, and
\begin{equation}
D_\rho(q)=m^2_\rho-q^2-i\sqrt{q^2}\Gamma_{\rho\pi\pi}(q^2)
\label{rhoprop}
\end{equation}
is the inverse propagator of $\rho$ meson whose width is dominated by the
$\pi\pi$ decay mode:
\begin{equation}
\Gamma_{\rho\pi\pi}(q^2)=\Gamma_{\rho\pi\pi}(m^2_\rho)
{m^2_\rho\over q^2}\left({q^2-4m^2_\pi\over m^2_\rho-4m^2_\pi}
\right)^{3/2}.
\label{rhowidth}
\end{equation}

Yet even in this simplified form the expressions for the $\omega\to5\pi$
amplitudes are not easy to use for evaluation of the branching ratios. To
go further, it should be noted the following.
One can check that the invariant mass of the $4\pi$ system on which the
contribution of the diagrams shown in Fig.~\ref{fig1}(a) depends, changes in
very narrow range $558\mbox{ MeV}<m_{4\pi}<642\mbox{ MeV}$. Hence,
one can set it in all the $\rho$ propagators standing as the first terms in
all square brackets in Eqs.~(\ref{om3pi0}) and (\ref{om1pi0}),
with the accuracy $20\%$ in width,
to the equilibrium value ${\overline{m^2_{4\pi}}}^{1/2}=620$ MeV evaluated
for the pion energy $E_\pi=m_\omega/5$ which gives the dominant
contribution. The same is true for the invariant mass of the pion pairs on
which the $\rho$ propagators standing as the last terms in square brackets
of the above expressions depend. This invariant mass varies in the
narrow range $280\mbox{ MeV}<m_{2\pi}<360\mbox{ MeV}$. With the same accuracy,
one can set it to $\overline{m^2_{2\pi}}^{1/2}=295$ MeV in all relevant
propagators. On the other hand, the amplitude of the process
$\omega\to\rho^0\pi^0\to(2\pi^+2\pi^-)\pi^0$ is
\begin{eqnarray}
M[\omega\to\rho^0\pi^0\to(2\pi^+2\pi^-)\pi^0]
&=&4{N_cg_{\rho\pi\pi}g^2\over8(2\pi)^2f^3_\pi}
\nonumber\\
&&\times
\varepsilon_{\mu\nu\lambda\sigma}q_\mu\epsilon_\nu(q_1+q_2)_\lambda
\nonumber\\
&&\times{q_{4\sigma}\over D_\rho(q-q_4)},
\label{ampli}
\end{eqnarray}
where the momentum assignment is the same as in
Eq.~(\ref{om1pi0}). The other relevant amplitude corresponding to
the first diagram in Fig.~\ref{fig1}(b) is
\widetext
\begin{eqnarray}
M[\omega\to\rho^0\pi^0\to(\pi^+\pi^-)(\pi^+\pi^-\pi^0)]&=&
{N_cg_{\rho\pi\pi}g^2\over8(2\pi)^2f^3_\pi}
\varepsilon_{\mu\nu\lambda\sigma}q_\mu\epsilon_\nu
\nonumber\\
&&\times(1+P_{12})(1+P_{35}){q_{1\lambda}q_{3\sigma}\over D_\rho(q_1+q_3)}
\label{ampli1}
\end{eqnarray}
\narrowtext Then, taking into account the above consideration
concerning the invariant masses, and comparing
Eqs.~(\ref{om1pi0}), (\ref{ampli}), and (\ref{ampli1}), one can
see that
\begin{eqnarray}
M(\omega\to2\pi^+2\pi^-\pi^0)&\approx&{5\over2}
M[\omega\to\rho^0\pi^0\to(2\pi^+2\pi^-)\pi^0]   \nonumber\\
&&\times\left[1-{D_\rho(\overline{m^2_{4\pi}})
\over2D_\rho(\overline{m^2_{2\pi}})}\right].
\label{ampli2}
\end{eqnarray}
The same treatment shows that
\begin{eqnarray}
M(\omega\to\pi^+\pi^-3\pi^0)&\approx&{5\over2}
M[\omega\to\rho^+\pi^-\to(\pi^+3\pi^0)\pi^-]   \nonumber\\
&&\times\left[1-{D_\rho(\overline{m^2_{4\pi}})
\over2D_\rho(\overline{m^2_{2\pi}})}\right],
\label{ampli3}
\end{eqnarray}
where
\begin{eqnarray}
M[\omega\to\rho^+\pi^-\to(\pi^+3\pi^0)\pi^-]
&=&-4{N_cg_{\rho\pi\pi}g^2\over8(2\pi)^2f^3_\pi}
\nonumber\\
&&\times
{\varepsilon_{\mu\nu\lambda\sigma}q_\mu\epsilon_\nu q_{1\lambda}q_{2\sigma}
\over D_\rho(q-q_2)},
\label{ampli4}
\end{eqnarray}
and the final momenta assignment is the same as in Eq.~(\ref{om3pi0}).
The numerical values of $\overline{m^2_{4\pi}}^{1/2}$ and
$\overline{m^2_{2\pi}}^{1/2}$
found above are such that  the correction factor in parentheses of
Eqs.~(\ref{ampli1}) and (\ref{ampli2}) amounts to $20\%$ in magnitude.
In what follows, the above correction will be taken into account as an
overall factor of 0.64 in front of the branching ratios of the decays
$\omega\to5\pi$. When making this estimate, the imaginary part of the $\rho$
propagators in square brackets of Eq.~(\ref{ampli1}) and (\ref{ampli2})
is neglected. This assumption is valid with the accuracy better than
$1\%$ in width.

It would be useful to fulfill the model estimate of
the $\omega\to5\pi$ partial widths as follows.
The corresponding equilibrium
pion momenta are $|{\bf q}_{\pi^+}|=70$ MeV and
$|{\bf q}_{\pi^0}|=79$ MeV. The integrations over angles of final pions
can be fulfilled assuming them independent. Using the nonrelativistic
expression for phase space of five pions \cite{byck},
\begin{equation}
R_5={\pi^6(\prod^5_im_{\pi i})^{1/2}\over60(\sum^5_im_{\pi i})^{3/2}}
(m_\omega-\sum^5_im_{\pi i})^5
\label{r5}
\end{equation}
whose numerical value coincides with the accuracy  $1\%$
with the numerically evaluated exact expression,
and introducing the branching ratio
at the $\omega$ mass as
\begin{equation}
B_{\omega\to5\pi}=\Gamma_{\omega\to5\pi}(m_\omega)/\Gamma_\omega,
\label{brom}
\end{equation}
one finds
\begin{eqnarray}
B(\omega\to5\pi)&\simeq&\left[{5N_c\over2\pi^2}\left({g_{\rho\pi\pi}g^2
\over8f^3_\pi}\right){m_\omega|{\bf q}_{\pi^+}|
\over |D_\rho(\overline{m^2_{4\pi}})|}\right]^2     \nonumber\\
&&\times{R_5\over18(2\pi)^{11}m_\omega\Gamma_\omega}
\left|1-{D_\rho(\overline{m^2_{4\pi}})
\over2D_\rho(\overline{m^2_{2\pi}})}\right|^2   \nonumber\\
&&\times\left\{|{\bf q}_{\pi^0}|^2 (2\pi^+2\pi^-\pi^0)\atop
{|{\bf q}_{\pi^+}|^2\over3} (\pi^+\pi^-3\pi^0)\right.
\label{bom5pi}
\end{eqnarray}
The calculation  gives
$B( \omega\to2\pi^+2\pi^-\pi^0)=2.5\times10^{-9}$ and
$B(\omega\to\pi^+\pi^-3\pi^0)=1.0\times10^{-9}$.

The evaluation of the partial widths valid with accuracy $20\%$
can be obtained upon taking the amplitude of each considered decays as
5/2 times the $\rho\pi$ production state amplitude with the subsequent
decay $\rho\to4\pi$, and calculate the partial width
using the calculated widths of the latter:
\begin{eqnarray}
B_{\omega\to2\pi^+2\pi^-\pi^0}&=&\left|1-{D_\rho(\overline{m^2_{4\pi}})
\over2D_\rho(\overline{m^2_{2\pi}})}\right|^2\left({5\over2}\right)^2
{2\over\pi\Gamma_\omega}  \nonumber\\
&&\times\int_{4m_{\pi^+}}^{m_\omega-m_{\pi^0}}
dm   \nonumber\\
&&\times{m^2\Gamma_{\omega\to\rho^0\pi^0}(m)
\Gamma_{\rho\to2\pi^+2\pi^-}(m)\over|D_\rho(m^2)|^2}
\nonumber\\
&&=1.1\times10^{-9}
\label{b1pi0}
\end{eqnarray}
where $$\Gamma_{\omega\to\rho^0\pi^0}(m)=g^2_{\omega\rho\pi}q^3
(m_\omega,m,m_{\pi^0})/12\pi,$$
$$g_{\omega\rho\pi}={N_cg^2\over8\pi^2f_\pi}=14.3\mbox{ GeV}^{-1}.$$
Note also the $a^{-1}$ dependence of the $\omega\to5\pi$ width on the HLS
parameter $a$. The branching ratio
$B_{\omega\to\pi^+\pi^-3\pi^0}$
is obtained from Eq.~(\ref{b1pi0}) upon inserting the lower integration
limit to $m_{\pi^+}+3m_{\pi^0}$,  $m_{\pi^0}\to m_{\pi^+}$
in the expression for the momentum $q$ and substitution
of the $\rho^+\to\pi^+3\pi^0$ decay width corrected for the mass
difference of charged and neutral pions. Of course, the main correction
of this sort comes from the phase space volume of the final 4$\pi$ state.
One obtains
\begin{eqnarray}
B_{\omega\to\pi^+\pi^-3\pi^0}&=&\left|1-{D_\rho(\overline{m^2_{4\pi}})
\over2D_\rho(\overline{m^2_{2\pi}})}\right|^2\left({5\over2}\right)^2
{2\over\pi\Gamma_\omega}  \nonumber\\
&&\times\int_{m_{\pi^+}+3m_{\pi^0}}^{m_\omega-m_{\pi^+}}
dm   \nonumber\\
&&\times{m^2\Gamma_{\omega\to\rho^+\pi^-}(m)
\Gamma_{\rho^+\to\pi^+3\pi^0}(m)\over|D_\rho(m^2)|^2}
\nonumber\\
&&=8.5\times10^{-10}
\label{b1pi01}
\end{eqnarray}
where $$\Gamma_{\omega\to\rho^+\pi^-}(m)=g^2_{\omega\rho\pi}q^3
(m_\omega,m,m_{\pi^+})/12\pi.$$
As is pointed out in Ref.~\cite{bando}, the inclusion of the direct
$\omega\to\pi^+\pi^-\pi^0$ vertex reduces the 3$\pi$ decay width of the
$\omega$ by $33\%$. This implies that one should make the following
replacement to take into account the effect of  the pointlike diagrams in
Fig.~\ref{fig1}(b) in the expression for the suppression factor:
\begin{eqnarray}
\left|1-{D_\rho(\overline{m^2_{4\pi}})
\over2D_\rho(\overline{m^2_{2\pi}})}\right|^2&\to&
\left|1-{D_\rho(\overline{m^2_{4\pi}})\over2}\left[{1\over
D_\rho(\overline{m^2_{2\pi}})}\right.\right.  \nonumber\\
&&\left.\left.-{1\over3m^2_\rho}\right]\right|^2
\approx\left|1-{D_\rho(\overline{m^2_{4\pi}})
\over3D_\rho(\overline{m^2_{2\pi}})}\right|^2    \nonumber\\
&&\simeq0.75,
\label{correc}
\end{eqnarray}
\narrowtext
instead of 0.64, which results in the increase of the above branching ratios
by the factor of 1.17.

\begin{figure}
\centerline {\epsfysize=7in \epsfbox{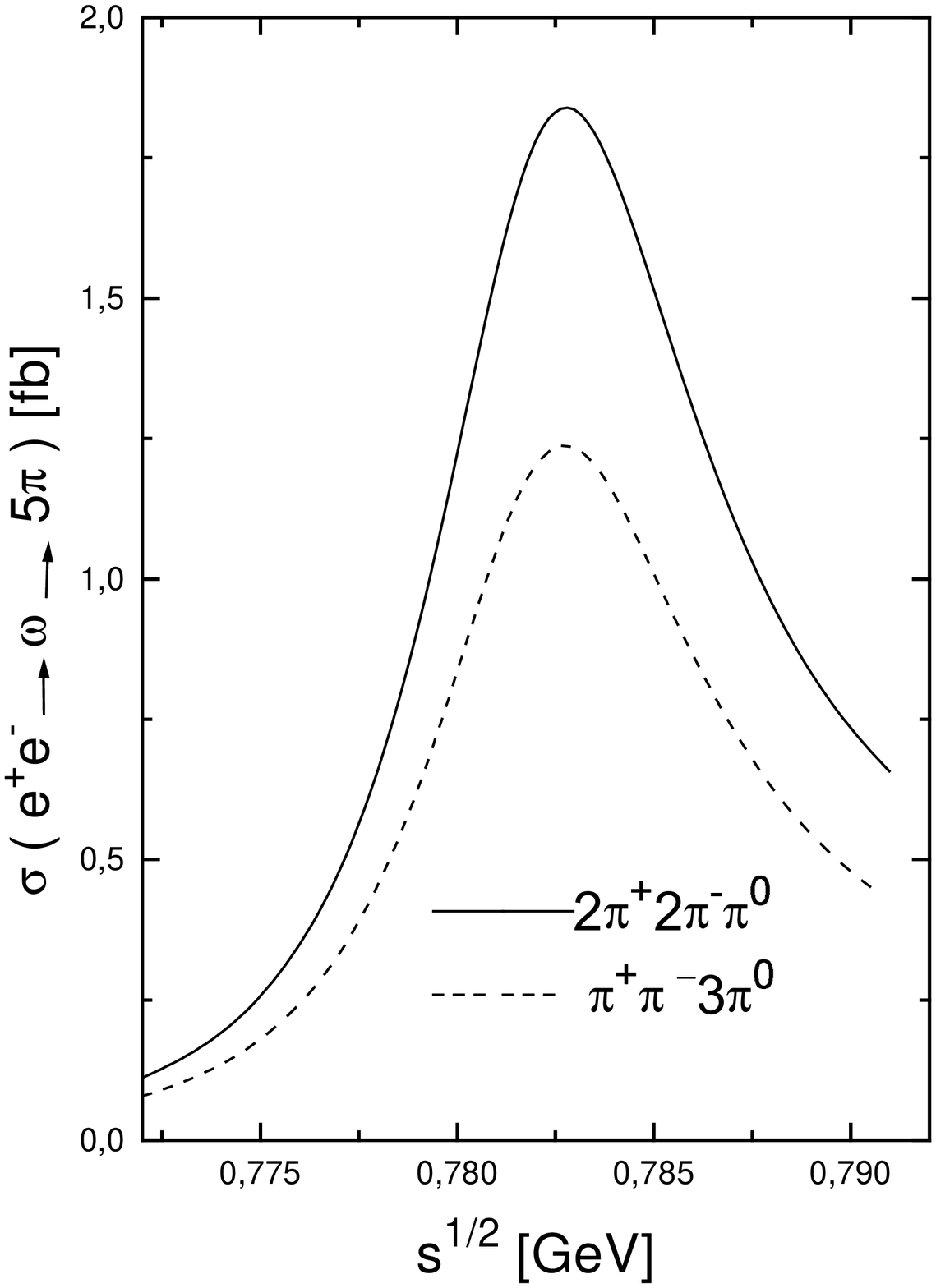}}
\caption{The $\omega\to5\pi$ excitation curves in
$e^+e^-$  annihilation in the vicinity of the $\omega$ resonance.
\label{fig2}}
\end{figure}

The numerical value of the $\omega\to5\pi$ decay width changes by the factor
of two when varying the energy within $\pm\Gamma_\omega/2$
around the $\omega$
mass. In other words, the dependence of
this partial width on energy is very strong.
This is illustrated by Fig.~\ref{fig2} where the
$\omega\to5\pi$ excitation curves in $e^+e^-$ annihilation,
\begin{eqnarray}
\sigma_{e^+e^-\to\omega\to5\pi}(s)&=&12\pi\left({m_\omega\over E}\right)^3
\Gamma_{\omega e^+e^-}(m_\omega)   \nonumber\\
&&\times{\Gamma_\omega B_{\omega\to5\pi}(E)\over\left[(s-m^2_\omega)^2+
(m_\omega\Gamma_\omega)^2\right]},
\label{omee}
\end{eqnarray}
are plotted. Here $B_{\omega\to2\pi^+2\pi^-\pi^0}(E)$
[$B_{\omega\to\pi^+\pi^-3\pi^0}(E)$] is given by Eq.~(\ref{b1pi0})
[(\ref{b1pi01})], respectively, with the substitution $m_\omega\to
E$. The mentioned strong energy dependence of the partial width
results in the asymmetric shape of the $\omega$ resonance and the
shift of its peak by $+0.7$ MeV . As is seen from Fig.~\ref{fig2},
the peak value of the $5\pi$ state production cross section is
about 1.5-2.0 femtobarns. Yet the decays $\omega\to5\pi$ can be
observable on $e^+e^-$ colliders. Indeed, with  the luminosity
$L=10^{33}\mbox{cm}^{-2}\mbox{s}^{-1}$ near the $\omega$ peak,
which seems to be feasible,  one may expect about 2 events per
week for the considered decays to be detected at these colliders.

As for the angular distributions of the final pions  are concerned, they,
of course, should be deduced from the full amplitudes Eqs.~(\ref{om1pi0})
and (\ref{om3pi0}). However, some qualitative
conclusions about the angular distributions
can be drawn from the simplified expressions
Eqs.~(\ref{ampli}), (\ref{ampli2}). Since  helicity is conserved,
only the states of the
$\omega(782)$ with the spin projections $\lambda=\pm1$ on the
$e^+e^-$ beam axes given by the unit vector ${\bf n}_0$ are
populated.
In what follows the suitable notation for the vector product of the pion
momenta are used:
\begin{eqnarray}
[{\bf q}_i\times{\bf q}_j]&=&|{\bf q}_i||{\bf q}_j|\sin\theta_{ij}
\nonumber\\
&&\times(\sin\Theta_{ij}\cos\Phi_{ij},\sin\Theta_{ij}\sin\Phi_{ij},
\cos\Theta_{ij}).
\label{not}
\end{eqnarray}
In other words, $\theta_{ij}$ is the angle between the pion momenta
${\bf q}_i$ and ${\bf q}_j$, $\Theta_{ij}$, $\Phi_{ij}$ being the polar and
azimuthal angles of the normal to the plane spanned by the  momenta
${\bf q}_i$ and ${\bf q}_j$.
Choosing ${\bf n}_0$ to be the unit vector along z axes,
the probability density of the emission of two $\pi^+$'s with the momenta
${\bf q}_1$, ${\bf q}_2$, and $\pi^0$ with the momentum ${\bf q}_4$
is represented as
\begin{eqnarray}
w&\propto&\left[{\bf q}_4\times\left({\bf q}_1+{\bf
q}_2\right)\right]^2-\left({\bf n}_0\cdot\left[{\bf
q}_4\times\left({\bf q}_1+{\bf q}_2\right)\right]\right)^2
\nonumber\\
&&={\bf q}^2_4\left[{\bf q}^2_1\sin^2\theta_{41}\sin^2\Theta_{41}+
{\bf q}^2_2\sin^2\theta_{42}\sin^2\Theta_{42}\right. \nonumber\\
&&\left.+2|{\bf q}_1||{\bf q}_2|\sin\Theta_{41}\sin\Theta_{42}
\sin\theta_{41}\sin\theta_{42}\right.\nonumber\\
&&\left.\times\cos(\Phi_{41}-\Phi_{42})\right]
\label{anc1}
\end{eqnarray}
in the case of the final state $2\pi^+2\pi^-\pi^0$. Here the
momentum assignment is the same as in Eq.~(\ref{om1pi0}).
The angular distribution of two $\pi^-$'s with the momenta ${\bf q}_3$,
${\bf q}_5$, and $\pi^0$
is obtained from Eq.~(\ref{anc1}) upon the replacement
${\bf q}_{1,2}\to{\bf q}_{3,5}$, because the identity
$\varepsilon_{\mu\nu\lambda\sigma}q_\mu\epsilon_\nu(q_1+q_2)_\lambda
q_{4\sigma}=-\varepsilon_{\mu\nu\lambda\sigma}q_\mu\epsilon_\nu
(q_3+q_5)_\lambda q_{4\sigma}$ is valid.
Since another identity
$\varepsilon_{\mu\nu\lambda\sigma}q_\mu\epsilon_\nu(q_1+q_2)_\lambda
q_{4\sigma}=-\varepsilon_{\mu\nu\lambda\sigma}q_\mu\epsilon_\nu
(q_1+q_2)_\lambda(q_3+q_5)_\sigma$
is valid, one can write the angular distribution that includes four charged
pions:
\widetext
\begin{eqnarray}
w&\propto&[({\bf q}_1+{\bf q}_2)\times({\bf q}_3+{\bf q}_5)]^2
-\left({\bf n}_0\cdot
[({\bf q}_1+{\bf q}_2)\times({\bf q}_3+{\bf q}_5)]\right)^2
\nonumber\\
&&=(1+P_{12})(1+P_{35}){\bf q}^2_1{\bf q}^2_3
\sin^2\theta_{13}\sin^2\Theta_{13}\nonumber\\
&&+2|{\bf q}_1||{\bf q}_2|(1+P_{35}){\bf q}_3^2\sin\theta_{13}\sin\theta_{23}
\sin\Theta_{13}\sin\Theta_{23}\cos(\Phi_{13}-\Phi_{23})
\nonumber\\
&&+2|{\bf q}_3||{\bf q}_5|(1+P_{12}){\bf q}_1^2\sin\theta_{13}\sin\theta_{15}
\sin\Theta_{13}\sin\Theta_{15}\cos(\Phi_{13}-\Phi_{15})
\nonumber\\
&&+2|{\bf q}_1||{\bf q}_2||{\bf q}_3||{\bf q}_5|(1+P_{35})
\sin\theta_{13}\sin\theta_{25}\sin\Theta_{13}\sin\Theta_{25}
\cos(\Phi_{13}-\Phi_{25}).
\label{anc2}
\end{eqnarray}
\narrowtext Here $P_{ij}$  interchanges the indices $i$ and $j$.
In the case of the final state $\pi^+\pi^-3\pi^0$ the
corresponding probability density can be obtained from
Eqs.~(\ref{ampli3}) and (\ref{ampli4}) and looks as
\begin{eqnarray}
w&\propto&[{\bf q}_1\times{\bf q}_2]^2-({\bf n}_0\cdot[{\bf
q}_1\times{\bf q}_2])^2       \nonumber\\
&&={\bf q}^2_1{\bf q}^2_2\sin^2\theta_{21}\sin^2\Theta_{21}.
\label{ann1}
\end{eqnarray}
Here the momentum assignment is the same as in Eq.~(\ref{om3pi0}).
The corresponding angular distribution of one charged, say
$\pi^+$, and three neutral pions can be obtained from Eqs.~(\ref{ampli3})
and (\ref{ampli4}) upon using the identity
$\varepsilon_{\mu\nu\lambda\sigma}q_\mu\epsilon_\nu q_{1\lambda}q_{2\sigma}
=-\varepsilon_{\mu\nu\lambda\sigma}q_\mu\epsilon_\nu
q_{1\lambda}(q_3+q_4+q_5)_\sigma$ and looks as
\begin{eqnarray}
w&\propto&\left[{\bf q}_1\times\sum_i{\bf q}_i\right]^2
-\left({\bf n}_0\cdot\left[{\bf q}_1\times\sum_i{\bf q}_i\right]\right)^2
\nonumber\\
&&={\bf q}^2_1\left[\sum_i{\bf q}^2_i\sin^2\theta_{i1}\sin^2\Theta_{i1}
\right. \nonumber\\
&&\left.+2\sum_{i\not=j}|{\bf q}_i||{\bf q}_j|\sin\theta_{i1}\sin\theta_{j1}
\sin\Theta_{i1}\sin\Theta_{j1}\right.\nonumber\\
&&\left.\times\cos(\Phi_{i1}-\Phi_{j1})\right].
\label{ann2}
\end{eqnarray}
Here indeces $i,j$ run over 3,4,5.

The strong energy dependence of the five pion partial width of the $\omega$
implies that the
branching ratio at the $\omega$ mass, Eq.~(\ref{brom}),
evaluated above, is slightly different from that determined by
the expression
\begin{equation}
B^{\rm aver}_{\omega\to5\pi}(E_1,E_2)=
{2\over\pi}\int_{E_1}^{E_2}dE{E^2\Gamma_\omega B_{\omega\to5\pi}(E)\over
(E^2-m^2_\omega)^2+(m_\omega\Gamma_\omega)^2}.
\label{bmom}
\end{equation}
Taking $E_1=$ 772 MeV and $E_2=$ 792 MeV, one finds
$B^{\rm aver}_{\omega\to2\pi^+2\pi^-\pi^0}(E_1,E_2)=9.0\times10^{-10}$ and
$B^{\rm aver}_{\omega\to\pi^+\pi^-3\pi^0}(E_1,E_2)=6.7\times10^{-10}$
to be compared
to Eq.~(\ref{b1pi0}) and (\ref{b1pi01}), respectively. In particular,
the quantity $B^{\rm aver}_{\omega\to2\pi^+2\pi^-\pi^0}(E_1,E_2)$ is the
relevant characteristics  of this specific decay
mode in photoproduction experiments.
The Jefferson Lab "photon factory" \cite{dzierba}
could also be suitable for detecting the
five pion decays of the $\omega$. However, in view of the suppression of the
$\omega$ photoproduction cross section by the factor of 1/9 as compared with
the $\rho$ one, the total number of $\omega$ mesons will amount to
$7\times10^8$ per nucleon. Hence, the increase of intensity of this machine
by the factor of 50 is highly desirable, in order to observe the decay
$\omega\to5\pi$ and measure its branching ratio.
Evidently, the $\omega$ photoproduction
on heavy nuclei is preferable in view of the dependence of the cross
section on atomic weight $A$ growing as $A^{0.8-0.95}$ \cite{leith}.

The conclusions about the angular distributions in photoproduction
are the following.
Of course, their  general expression should be deduced from
the full decay amplitudes Eqs.~(\ref{om1pi0}) and (\ref{om3pi0}), together
with the detailed form of the photoproduction mechanism. The qualitative
picture, however, can be obtained upon noting that $s$-channel helicity
conservation is a good selection rule for the photoproduction reactions. Then
in the helicity reference frame characterized as the frame where the $\omega$
is at rest, while its spin quantization axes is directed along the $\omega$
momentum in the center-of-mass system, the expressions for the angular
distributions coincide with the expressions Eqs.~(\ref{anc1}),(\ref{anc2}),
(\ref{ann1}) and (\ref{ann2}).
Since, at high energies,  the direction of the final
$\omega$ momentum lies at the scattering angle less than $0.5^\circ$
in the case of the photoproduction on heavy nuclei,
the vector ${\bf n}_0$ can be treated as pointed along
the photon beam direction.

Together with the $e^+e^-$ annihilation experiments, the study of
the photoproduction of the five pion states on heavy nuclei would
also allow to measure the corresponding partial width of the
$\omega(782)$. The comparison with theoretical expectations
presented here would give the possibility of testing the
predictions of chiral models that include the vector mesons in the
situation where the decay amplitudes are determined by very low
pion momenta.

We are grateful to G.~N.~Shestakov for discussion. The present work is
supported in part by the grant RFBR-INTAS IR-97-232.

\end{document}